\documentclass[%
 aip,
 amsmath,amssymb,nofootinbib,floatfix,
 reprint,%
]{revtex4-1}

\usepackage{graphicx}% Include figure files
\usepackage{dcolumn}% Align table columns on a decimal point

\usepackage{verbatim}
\usepackage{lineno}
\usepackage{xspace} 
\usepackage[utf8]{inputenc}
\usepackage[T1]{fontenc}
\usepackage{mathptmx}
\usepackage{etoolbox}
\usepackage{comment}
\usepackage{xcolor}
\usepackage{graphicx}% Include figure files
\usepackage{dcolumn}% Align table columns on decimal point
\usepackage{bm}% bold math
\newcommand{\axpwr}{P_{\rm ax}\xspace}

\makeatletter
\def\@email#1#2{%
 \endgroup
 \patchcmd{\titleblock@produce}
  {\frontmatter@RRAPformat}
  {\frontmatter@RRAPformat{\produce@RRAP{*#1\href{mailto:#2}{#2}}}\frontmatter@RRAPformat}
 {}{}
}%
\makeatother

\begin{document}

\preprint{AIP/123-QED}

\title[Dielectric elements in axion haloscopes]{On the use of dielectric elements in axion searches with microwave resonant cavities}

\author{Xiran Bai}%
 \email{xiran.bai@yale.edu.}
\affiliation{ 
Department of Physics, Yale University, New Haven, Connecticut 06520, USA%\\This line break forced with \textbackslash\textbackslash
}%
\affiliation{
Wright Laboratory, Department of Physics, Yale University, New Haven, Connecticut 06520, USA
}%

\author{Michael J. Jewell}
\affiliation{ 
Department of Physics, Yale University, New Haven, Connecticut 06520, USA%\\This line break forced with \textbackslash\textbackslash
}%
\affiliation{
Wright Laboratory, Department of Physics, Yale University, New Haven, Connecticut 06520, USA
}%

\author{Steve K. Lamoreaux}%
\affiliation{ 
Department of Physics, Yale University, New Haven, Connecticut 06520, USA%\\This line break forced with \textbackslash\textbackslash
}%
\affiliation{
Wright Laboratory, Department of Physics, Yale University, New Haven, Connecticut 06520, USA
}%

\author{Reina H. Maruyama}
\affiliation{ 
Department of Physics, Yale University, New Haven, Connecticut 06520, USA%\\This line break forced with \textbackslash\textbackslash
}%
\affiliation{
Wright Laboratory, Department of Physics, Yale University, New Haven, Connecticut 06520, USA
}%

\author{Karl van Bibber}%
\affiliation{ 
Department of Nuclear Engineering, University of California, Berkeley, California 94720, USA
}%

\date{04/13/2023}% It is always \today, today,
             %  but any date may be explicitly specified

\begin{abstract}
This study explores the primary effects of dielectric materials in a resonant cavity-based search for axion dark matter. While dielectrics prove beneficial in numerous cases, their incorporation may lead to less-than-optimal performance, especially for the lowest TM mode. Additionally, the stronger confinement of the electric field inside the dielectrics can exacerbate mode mixings, in particular for higher-order modes. Case studies have been carried out using a combination of analytical solutions and numerical simulations. The findings indicate dielectric cavities employing the $\text{TM}_{010}$ mode experience a significant reduction in sensitivity when compared to a similar search conducted in a cavity at equivalent frequency using no dielectrics. 
\end{abstract}

\maketitle

\section{\label{intro}Introduction}
 By introducing a dynamic phase term in the QCD Lagrangian, Peccei and Quinn provided an explanation of the Charge-Parity conservation in strong interactions\cite{peccei1977CP,peccei1977CP2}, which was further interpreted by Wilczek as implying a new pseudoscalar particle, the axion.  The possibility of an axion mass at the weak scale was quickly experimentally ruled out, however there is no other direct standard model constraint on the lower mass limit. Sufficiently light axions are a natural candidate for dark matter, as originally proposed by Sikivie \cite{axionDM_1983}. To detect dark matter axions, Sikivie proposed the haloscope technique, in which axions are coupled to the electromagnetic fields in a resonant cavity, through the inverse Primakoff process due to the application of a strong static magnetic field \cite{Sikivie:1983ip_halotheory}. The frequency of the oscillating field is given by the axion mass $m_a$ as $\nu=m_a c^2/h$. The on-resonance axion conversion power in such a cavity is proportional to
\begin{equation}
    \axpwr \propto g^{2}_{a\gamma\gamma}\frac{\rho_a}{m_a} B^{2}_{0} V C_{mnl} Q_{0}
    \label{eq-axion_power}
\end{equation}
where $g_{a\gamma\gamma}$ is the axion-photon coupling constant, $\rho_a$ is the local axion mass density, $B_0$ is the strength of the static magnetic field, $V$ is the effective volume of the cavity, and $Q_0$ is the unloaded cavity quality factor. $C_{mnl}$ is the normalized form factor describing the coupling of the axion to a specific mode labeled by $mnl$, and is given by
\begin{equation}
    C_{mnl} = \frac{|\int_{V} \boldsymbol{B_0}(\boldsymbol{x}) \cdot \boldsymbol{E}( \boldsymbol{x}) \,dx^3|^2}
    {B_0^2V \int_{V} \epsilon_0 \epsilon_d(\boldsymbol{x}) | \boldsymbol{E}(\boldsymbol{x})|^2  \,dx^3}.
    \label{eq:formfac}
\end{equation}
where $\boldsymbol{E}(\boldsymbol{x})$ is the oscillating electric field vector amplitude of the particular mode,
$\epsilon_0$ is the permittivity of free space, and $\epsilon_d(\boldsymbol{x})$ represents the spatial dependence of the relative permittivity, which we will take as lumped elements with constant properties. 

Because the axion's mass is \textit{a priori} unknown, a search for galactic halo axion dark matter requires tuning the cavity to different resonant frequencies. To achieve some specified sensitivity to $g_{a\gamma\gamma}$ over some range of frequencies, the scan rate is proportional to \begin{equation}
    \frac{d\nu}{dt} \propto V^{2} C_{mnl}^{2} Q_{0}.
    \label{eq-scan_rate}
\end{equation}
Because the goal is to scan as quickly as possible while attaining a useful sensitivity to $g_{a\gamma\gamma}$, we define the frequency-dependent figure of merit for a cavity design, 
\begin{equation}
    \mathcal{F}(\nu) = V^2 C_{mnl}^2 Q_{0}\equiv \mathcal{F}
    \label{eq:FOM_scanrate}
\end{equation}
which needs to be maximized for an optimal search.

Recently, there has been a growing interest in the axion as a dark matter candidate, and a number of axion experiments represent the exploration of the use of dielectric materials\cite{morris_1984,CAPP,Alesini_2020_drod,QUAX_2022,QUAX_2021,ORGAN_2018}. Compared to metallic tuning elements, which in a practical sense change the internal cavity dimension and hence its resonant frequency, dielectric tuning elements allow the experimental search to lower frequency axions with the same cavity. Furthermore, if placed strategically, dielectric materials can increase the quality factor by reducing the power loss along the cavity's metallic wall\cite{Alesini_2020_drod,blair1987}. Additionally, because TEM modes only exist in structures with a central conductor, replacing the metal tuning element with dielectrics can eliminate the TEM mode mixing\cite{simanovskaia_microwave_2023}. Furthermore, in searches utilizing higher-order modes, dielectric materials are used to improve the form factor by suppressing the opposite-phase electric field components\cite{Kim_2020,ORGAN_2018,QUAX_2022}. While dielectrics offer benefits in many cases, there are potential drawbacks to be considered. As we will show, introducing dielectrics causes energy concentration into the dielectrics, which may reduce the electric field that couples to the axion and exacerbate TE mode crossings. These factors need to be carefully taken into account when designing haloscope cavities.  

Although most of this knowledge had been established in the 1980s, the complete and full effects of dielectrics have not been coherently and simultaneously addressed together in the existing literature. In this study, we present an answer to a possible question, how much dielectric is too much? Pre-existing cavities can utilize dielectrics to move to a lower frequency range, but the impact of this has never been fully analyzed. Additionally, the role of the cavity is generally recognized as an impedance-matching system that, depending on the specific model, converts an "energy current" into electromagnetic field energy within the cavity; hence we want the highest impedance possible. This point has been recently studied \cite{Chaudhuri_2021} with a model slightly different from ours, however the full range of effect for cavity resonators was not addressed.  With this in mind, the inclusion of extra capacitance in a resonator tends to lower its impedance (see Appendix \ref{appen:circuit} for a lumped circuit model). We will explore the following effects due to the inclusion of dielectrics in a cavity:
\begin{enumerate}
\item The energy density in a dielectric-loaded cavity tends to reside in the $\boldsymbol{D}=\epsilon \boldsymbol{E}$ field, for a given cavity energy density, which is given by the volume integral of $\boldsymbol{D}\cdot \boldsymbol{E}$.  The effect is contained in Eq. (\ref{eq:formfac}), which tends to reduce the cavity form factor using the lowest TM mode.

\item The presence of dielectrics tends to reduce the electric field for a given cavity energy, and because the axion couples to the electric field, this directly reduces the rate of axion conversion to electromagnetic photons. This effect is implicit in the definition of the form factor.

\item The electromagnetic field dispersion relationship is modified (lower velocity of light) with dielectrics, which leads to a decrease in the resonant frequency for a fixed cavity volume or, conversely, a reduction in the cavity volume for a specific resonant frequency.

\item The inclusion of dielectrics can increase $Q_0$ by lowering the resistive loss at the cavity wall if placed deliberately. 

\item Spatially distributed dielectrics increase susceptibility to spurious TE modes. The stronger confinement of the electric fields inside the dielectric can exacerbate TE mode mixing, resulting in degraded axion sensitivity in those regions of the frequency space.    
\end{enumerate}

In the following, we will analyze the above effects, first separately then all together. The organization of this work is as follows. The discussion of a limiting case to illustrate the effect of the dielectric in a resonant cavity is presented in Sec.\ref{subsec:Dielectrics in Resonant Cavities}. Sec.\ref{sec:case_studies} explores more realistic use cases of dielectrics: the field solutions for a dielectric tuning rod cavity and a dielectric shell cavity are formulated and compared with a metal tuning rod cavity. The resulting comparison on the figure of merit is provided in Sec.\ref{sec:results}. Finally, design considerations for using dielectrics in haloscopes are summarized in Sec.\ref{disscussion and conclusion}.

\section{\label{subsec:Dielectrics in Resonant Cavities}The Effects of Dielectrics: a Limiting Case}

To understand how dielectrics in a haloscope impact the axion scan rate, one needs to consider the effects of dielectrics on $V$, $ C_{mnl}$, and $Q_0$ in Eq.\ref{eq:FOM_scanrate}. These effects are best illustrated by considering a simple limiting case where the entire cylindrical metal cavity is uniformly filled with dielectrics with a relative permittivity of $\epsilon_d > 1$. 

\begin{comment}
In this case, there is no tuning, and no need to consider intruder TE modes due to the cavity length that comes with a conductive tuning rod.

 Generally, with a metallic rod, the length of the cavity is roughly constrained to limit the number of TE mode crossings\cite{simanovskaia_microwave_2023}. As shown in Appendix \ref{appen:LAR}, this prevents the cavity from being made arbitrarily long because the mode density and the chances for mode crossings increase with the longitudinal aspect ratio, which is defined as the ratio of cavity length $L$ over the radius $\rho_c$.  However, we do not need to consider this for the present case but will return to it later.  We can assume the length is sufficiently long that the electric field only has an axial component.
 \end{comment}
 
\begin{comment}
   The resonant frequencies for the $\text{TM}_{lnm}$ and $\text{TE}_{lnm}$ modes are given by
$$\omega_{TM_{lnm}}=c\sqrt{\left(\frac{X_{ln}}{\rho_c}\right)^2+\left(\frac{m\pi}{L}\right)^2}$$
$$\omega_{TE_{lnm}}=c\sqrt{\left(\frac{X'_{ln}}{\rho_c}\right)^2+\left(\frac{m\pi}{L}\right)^2}$$
Where $X_{ln}$ are the zeros of Bessel function $J_{ln}(x)$ and $X'_{ln}$ are the zeros of $J'_{ln}(x)$, $\rho_c$ is the cavity radius, and $L$ is the cavity length.  
\end{comment}

 We will first consider the effect on cavity size by determining the cavity length and radius scaling required to keep the resonant frequency constant as the dielectric is added to the cavity. The cavity that contains dielectric will necessarily have a lower volume. The E-field of the $\text{TM}_{0n0}$ in a cylindrical cavity can be described by the Bessel function of the first kind $J_0(\rho)$. Assuming the gap between the dielectric and the cavity wall is small and almost negligible, the field is determined by the boundary condition that it vanishes outside the dielectric at the metal surface, namely $\bm{E}(\rho=\rho_c)=0$, where $\rho_c$ is the radius of the cavity, and is given by
\begin{equation}
    \rho_c = \frac{X_{0n}}{\sqrt{\epsilon_d}k_0}
    \label{eq:radius_d}
\end{equation}
where
\begin{equation}
k_0 = \sqrt{\mu_0\epsilon_0}\nu=\nu/c
\nonumber
\end{equation}
is the free-space wave number, with $\mu_0$ and $\epsilon_0$ the magnetic permeability and electric permittivity of the vacuum and $X_{0n}$ is the $n$th root of the zeroth order Bessel function.

To calculate how much volume is lost for the lowest TM mode in this limiting case, it can be seen from Eq.\ref{eq:radius_d} that the dielectric material reduces the radius by a factor of $\epsilon_d^{-1/2}$. While the length does not directly affect the frequency of the $\text{TM}_{010}$ mode, let us assume a finite cavity length L and scale it and the radius together. As shown in Appendix \ref{appen:LAR}, this restriction limits the number of mode crossings which increase with the aspect ratio $L/\rho_c$ and cause regions of insensitivity. Although there is no tuning in this limiting case, we adopt a $L/\rho_c$ as a general rule as it will be critical to later studies. The resulting reduction in volume is
\begin{equation}
\frac{\rho_{c}^2L}{\rho_0^2 L_0} \sim \epsilon_d^{-3/2}
\end{equation}
with $\rho_0$ and $L_0$ denoting the radius and length of an empty cavity with the same resonant frequency. Dielectric materials such as sapphire exhibit an extremely low $\tan\delta$ ($< 10^{-6}$) at cryogenic temperatures and are often used in dielectric resonators\cite{loss_tan_sapphire}. For a sapphire-filled cavity with $\epsilon_d$ of 10, the volume has a factor of $\sim 32$ reduction compared to the larger empty cavity operating at the same frequency.

Furthermore, a reduction in the form factor of the $\text{TM}_{010}$ mode occurs due to the electric field being reduced for a given energy density in the cavity. It is evident from Eq. (\ref{eq:formfac}) that the form factor of the dielectric-filled cavity is reduced by a factor $\epsilon_d$. In the case of sapphire, this results in a reduction in form factor by $10$.

Finally, to calculate the unloaded quality factor, we use 
\begin{equation}
    Q_0 = \omega\frac{U}{P_{\text{loss}}}
    \label{eq:quality_factor}
\end{equation}
which relates the energy stored in the cavity to the power loss ($P_{\text{loss}}$). For the purpose of this limiting case, we will ignore the dielectric loss from high-quality materials for now and only consider the metallic surface loss, which is given by 
\begin{equation}
    P_{\text{loss}} \approx P_{\text{surface}} = \frac{1}{2\sigma\delta} \int_{S}|H_{\parallel}|^2 \,d^2x
    \label{eq:P_loss_mwall}
\end{equation}
where $\sigma$ is the surface conductivity, $\delta$ is the skin depth, and $H_{\parallel}$ is the magnetic field parallel to the metal surfaces. Assuming the background electric field from axion conversion is represented by $\bm{E_0}$, we now calculate both the stored energy and the energy loss rate for that field.  As we have assumed the cavity length is very long, the energy density per unit length is approximately constant.  
% The $z$ component of the electric field is given by 
% \[ E_z=E_{0z} J_0({X_{01}\rho/\rho_c}\]
% as we have already discussed. 
To keep the frequency constant, the cavity length $L_d$ and radius $\rho_{c}$ each scale as $\epsilon_d^{-1/2}$ as already discussed.

From Maxwell's equations in the presence of dielectrics,
we have, for the dielectric-filled cavity,
\begin{equation}
\bm{\nabla}\times\bm{E}=-i\omega \bm{B}. 
\end{equation}
Due to the large extent of the cavity, we assume all components of the electric field are zero except for the z component given by $E_z(\rho)=E_{z0}J_0(X_{01}\rho/\rho_{c})$. Because $E_\parallel$ is continuous across the dielectric boundary this gives
\begin{equation}
\frac{i}{\omega}\bm{\nabla}\times\bm{E}=\frac{i}{\omega} E_{0z} \frac{X_{01}J_1(X_{01})}{\rho_{c}}= \bm{B}=\frac{1}{\mu_0} H_\parallel.
\end{equation} 
Thus, when $\rho=\rho_{c}$, $H_\parallel$ is constant, the surface integral from Eq. \ref{eq:P_loss_mwall} gives
\begin{equation}
\int_{S}|H_{\parallel}|^2 \,d^2x\propto
E_{0z}^2 2\pi\rho_{c}^{-1} L_d  \left(X_{01} J_1(X_{01})\right)^2.  
\end{equation}

$U$ is determined by the volume integral of $\bm{E}\cdot\bm{D}=\epsilon_d E_z^2$, (taking $\epsilon_d=1$ for the empty cavity) as
\begin{eqnarray}
U&=& \frac{\epsilon_0\epsilon_d}{2} \int_V E_{0z}^2 J_0(X_{01}\rho/\rho_{c})^2 2\pi \rho d\rho \nonumber\\
&=&\pi \epsilon_0\epsilon_d E_{0z}^2\pi L_d \rho_{c}^2 \int_0^1 J_0^2(X_{01}x)xdx.
\end{eqnarray}
We therefore find that 
\begin{equation}
Q_{0}=\omega \frac{U}{P_{surface}}\propto \epsilon_d \rho_{c}^3\propto \epsilon_d^{-1/2}. 
\end{equation}
which shows there is also a reduction in $Q_0$ which scales like $\epsilon_d^{-1/2}$. While dielectrics have the potential to increase the Q, it is not true in this limiting case because of the confined field in the dielectrics being too close to the cavity metal wall. 

\begin{comment}
axion conversion power from Eq.(\ref{eq-axion_power}) by a factor of 
\begin{equation}\label{axpower}
\axpwr\ \ (\propto C_{010} V Q_0) \rightarrow \epsilon_d^{-1}\epsilon_d^{-3/2}\epsilon_d^{-1/2}\axpwr \sim\epsilon_d^{-3}\axpwr \end{equation}
\end{comment}

Altogether, we find that in this limiting case, the addition of a dielectric dramatically suppresses the figure of merit $\mathcal{F}$ (hence also the scan rate) from Eq. (\ref{eq:FOM_scanrate}) by a factor of 

\begin{equation}
\label{eq:axscanrate}
%\frac{d\nu}{dt} \  (\propto C_{010}^2 V^2 %Q_0)\rightarrow \epsilon_d^{-2} \epsilon_d^{-3}
%\epsilon_d^{-1/2} \frac{d\nu}{dt}\sim\epsilon_d^{-11/2}\frac{d\nu}{dt}
\mathcal{F} \  (C_{010}^2 V^2 Q_0)\rightarrow \epsilon_d^{-2} \epsilon_d^{-3}
\epsilon_d^{-1/2}\mathcal{F}=\epsilon_d^{-11/2}\mathcal{F}
\end{equation}
indicating a significant reduction, with the loss increasing for dielectrics with larger $\epsilon_d$, which is a factor of $\sim316,000$ reduction in $\mathcal{F}$ if sapphire ($\epsilon_d\sim10$) is used as the filling material. While this simplified example is not meant as a realistic haloscope design, it demonstrates the main challenges which can arise when dielectric materials are used in cavity experiments. Alternatively, the reduction from dielectrics can be understood through impedance matching to the axion field. In this case, the addition of dielectrics makes the impedance matching between the cavity and the axion source less efficient as shown in Appendix \ref{appen:circuit}.  

%This result provides a striking and straightforward %reason why dielectric-loaded cavities are not %necessarily effective for haloscope resonators.

%While a completely dielectric-filled cavity has a %significant reduction in $\mathcal{F}$ for the lowest %TM mode, spatially distributed dielectrics can be %beneficial for higher-order modes. 

Having examined the lowest-order mode in a dielectric-filled cavity, let us briefly shift our focus to higher-order modes. In these cases, dielectrics can offer unique benefits: as mentioned in the previous section, the field concentration within dielectrics can augment the form factor by suppressing the opposite-phase electric field components typically associated with higher-order modes. However, higher-order modes are more susceptible to mode crossings with spurious TE modes, and the presence of dielectrics may intensify the issue through mode confinement and radius reduction. The problem of mode mixing further worsens for cavities with larger longitudinal aspect ratios (See Appendix \ref{appen:LAR} for more details). In this limiting case of a dielectric-filled cavity, this necessitates a longitudinal aspect ratio of:
\begin{equation}
\frac{L}{\rho_0} \lesssim \frac{\pi^2 Q_L}{600X_{0n}^3}\epsilon_0^{-1/2}.\\
\label{eq:AR_requirement_general}
\end{equation}
where the $Q_L$ is the loaded quality factor. A wave that has wavelength given by the cavity length $L$ would be the frequency of the first spurious mode and as the cavity is made longer, more enter the region of interest. This restriction is more stringent for higher-order modes as $L$ is inversely proportional to $X_{0n}$ which grows with increasing $n$. Additionally, for this limiting case, a radius reduction scaling with $\epsilon_0^{-1/2}$ further exacerbates the ratio, potentially resulting in a diminished detection volume.

In summary, in this limiting case, the performance of both the lowest-order mode and the higher-order modes are negatively affected by the presence of dielectric fillings. The former results in direct losses in signal power and scan rate, while the latter suffers from increased mode densities.

\begin{comment}
   One way to establish a reasonable ratio is to consider the cavity tuning more than 300 times its bandwidth before encountering a mode crossing, which is an achievable goal for a practical experiment that seeks to explore a broad range (See Appendix \ref{appen:LAR} for more details). This necessitates an longitudinal aspect ratio of: 
\end{comment}

%The general specifics of mode crowding depend on the %mode geometry and are beyond the scope of our current %discussion.
 
\begin{figure}
\includegraphics[width=\linewidth]{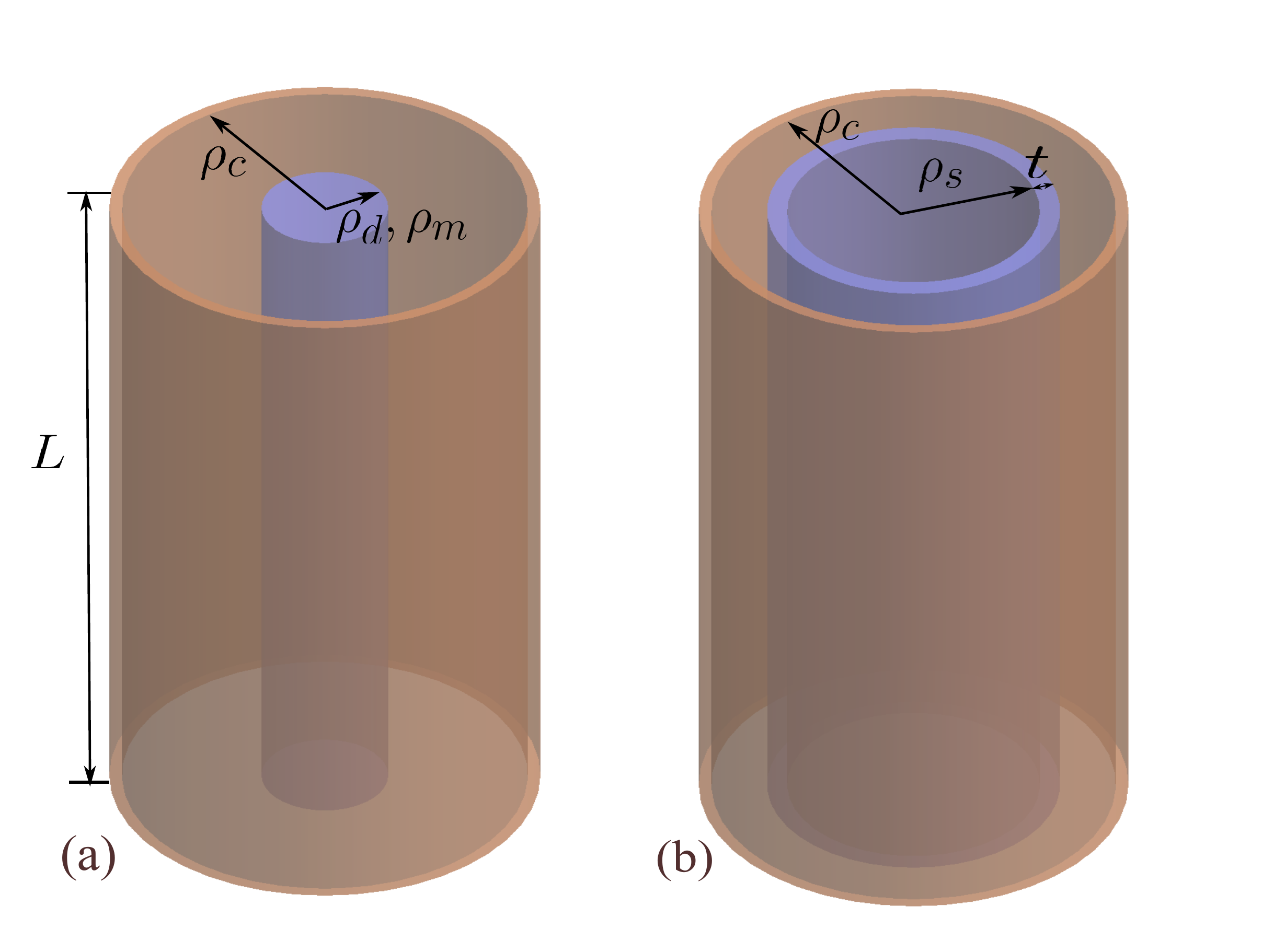}
%\captionsetup{justification=raggedright}
\caption{\label{fig:cavity_geometry} Geometries of partially filled cavities. (a): A tuning rod with radius $\rho_d$ (a dielectric rod) or $\rho_m$ (a metal rod) is placed in the center of the cavity with radius $\rho_c$. (b): A dielectric shell cavity, where a concentric shell of thickness $t$ is located inside the cavity at radius $\rho_s$. Both cavity walls are made of OFHC copper and their length over radius ratios are kept at $L/\rho_c=5$ as detailed in the Appendix \ref{appen:LAR}. The end caps for both cavities are not shown for simplicity.}
\end{figure}

\section{\label{sec:case_studies}Partially Filled Cavities}
While the uniformly filled cavity highlights the main effects of dielectric on the scan rate, a more careful study is needed to understand the impact on more realistic cavity designs. For more practical use of dielectrics in resonant cavities, we consider cavities that are partially filled with dielectric materials. This study will focus on the lowest-order TM mode of two cavity geometries that make use of dielectrics and we will compare their performance to that of a comparable search using a metal tuning rod. This section lays out the model for the three cases as well as the analytical field solutions used to compute $Q_0$, $C$, and $V$ for each. 

\begin{comment}
While the fundamental effects of introducing dielectrics remain the same, quantification of the impact for the more detailed geometry is harder and can only be computed analytically in certain configurations.
\end{comment}

\begin{figure}[ht]
    \centering
    \includegraphics[width=0.9\linewidth]{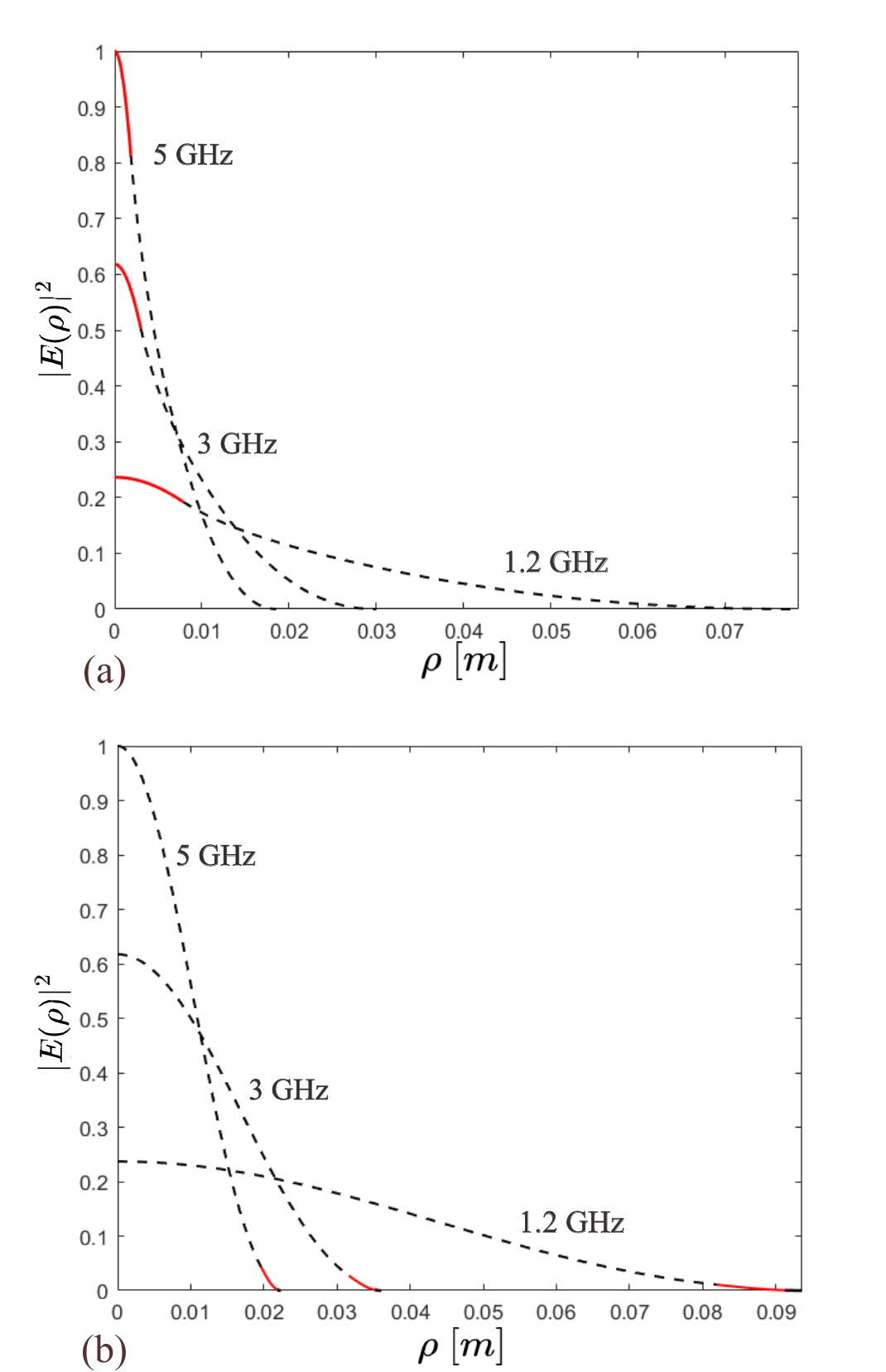}
%    \captionsetup{justification=raggedright}
    \caption{Example calculated E-fields as a function of the radial position within the cavity at different frequencies for (a) the dielectric tuning rod cavity, (b) the dielectric shell cavity. The red line indicates the dielectric region and the black dash line indicates the rest of the cavity (vacuum). The ratio of the rod and the shell thickness to the cavity radius ($\rho_d/\rho_c$ and $t/\rho_c$) are both 1/10, and the ratio of the shell position to the cavity radius ($\rho_s/\rho_c$) is 1/8, which are the optimal ratios selected in Sec.\ref{subsec:radialAR}.}
    \label{fig:field}
\end{figure}

\subsubsection{\label{subsec:MTRC}Metal Tuning Rod Cavity}
The benchmark design for comparison is a cylindrical metal cavity with a single metal tuning rod, which has been widely adopted in traditional haloscope experiments such as HAYSTAC. As shown in Fig.\ref{fig:cavity_geometry}(a), a metal tuning rod with radius $\rho_m$ is placed in the center of a cylindrical metal cavity with radius $\rho_c$. The geometry is simplified with the rod concentric with the cavity, allowing for the field solution to be found analytically and the gaps between the end caps and the rod are ignored. The metal used for both the cavity's main body and the rod is oxygen-free high thermal conductivity (OFHC) copper and the thickness is assumed to be larger than its skin depth.

For the $\text{TM}_{010}$ mode in a cylindrical cavity, the field solutions in the vacuum are described as a combination of the Bessel function of the first and second kind, $J_0$ and $N_0$ respectively, whereas the field inside the metal rod is 0. The electric fields in the longitudinal direction are thus given by

\begin{equation}
E(\rho) =
\begin{cases}
    0 & 0<\rho \leq \rho_m\\
    B_mJ_0(k_0\rho)+C_mN_0(k_0\rho) & \rho_m<\rho \leq \rho_c
\end{cases}
\label{eq:E_field_m}
\end{equation}
 Coefficients $B_m$ and $C_m$ are solved by enforcing boundary conditions that the field vanishes at the metal boundary, namely $E(\rho_m)=E(\rho_c)=0$. 

\begin{comment}
where $E_{m1}$ and $E_{m2}$ are the z component of the electric fields inside and outside the metal rod.
\begin{eqnarray}
\label{eq:E_field_m}
     \begin{array}{lcr}
    E_{m1}(\rho) = 0 & \mbox{}
    & 0<\rho \leq \rho_m\\
    E_{m2}(\rho) = B_mJ_0(k_0\rho)+C_mN_0(k_0\rho) & \mbox{} & \rho_m<\rho \leq \rho_c
\end{array}
\end{eqnarray} 
\end{comment}

The primary source of power loss in a metal rod cavity comes from the current induced by the magnetic field within the skin depth of the resistive metallic surfaces, including the cavity wall, the end caps, and the rod. Using Eqs.\ref{eq:E_field_m} and \ref{eq:P_loss_mwall} and the boundary condition,  we obtain the $Q_0$, $C_{010}$, and $V$ of the metal tuning rod cavity.

\subsubsection{\label{subsec:DTRC}Dielectric Tuning Rod Cavity}
One common use for dielectrics in haloscopes is to tune the resonant frequency. In this geometry, a solid dielectric tuning rod with radius $\rho_d$ is placed in the center of a cylindrical metal cavity with radius $\rho_c$ as shown in Fig.\ref{fig:cavity_geometry}(a). 

For the $\text{TM}_{010}$ mode, the electric fields in the longitudinal direction are again described by $J_0$ and $N_0$. Because $N_0$ diverges at 0, the field equations are therefore given by 
\begin{equation}
E(\rho) =
\begin{cases}
    A_dJ_0(\sqrt{\epsilon_d}k_0\rho) & 0<\rho \leq \rho_d\\
    B_dJ_0(k_0\rho)+C_dN_0(k_0\rho) & \rho_d<\rho \leq \rho_c
\end{cases}
\label{eq:E_field_d}
\end{equation}

\begin{comment}
where $E_{d1}$ and $E_{d2}$ are the electric fields inside and outside the dielectric.
  \begin{eqnarray}
\label{eq:E_field_d}
     \begin{array}{lcr}
    E_{d1}(\rho) = A_dJ_0(\sqrt{\epsilon_d}k_0\rho) & \mbox{}
    & 0<\rho \leq \rho_d\\
    E_{d2}(\rho) = B_dJ_0(k_0\rho)+C_dN_0(k_0\rho) & \mbox{} & \rho_d<\rho \leq \rho_c
\end{array}
\end{eqnarray}  
\end{comment}

\noindent where $A_d$, $B_d$, and $C_d$ are coefficients to be solved using the matching conditions at the boundary, which are given by
\begin{flalign}
E_{d1}(\rho_{d})&=E_{d2}(\rho_{d})\nonumber\\
\frac{\partial E_{d1}}{\partial \rho}|_{\rho_{d}}&=\frac{\partial E_{d2}}{\partial \rho}|_{\rho_{d}}\nonumber\\
E_{d2}(\rho_{c})&=0\nonumber
\end{flalign}

Although having a dielectric rod reduces the total metallic surface area, some additional power loss will come from the dielectric material itself. The dielectric loss is given by 
\begin{equation}
    P_{\text{volume}} = \frac{1}{2}\omega\epsilon_0\epsilon_d\tan{\delta}\int_{V}|E|^2 \,d^3x
    \label{eq:P_loss_d}
\end{equation}
where the $\tan{\delta}$ is the loss tangent of the dielectric material. In order to reduce the loss, the dielectric therefore needs to have a sufficiently low loss, such that the increased loss in the volume of the dielectric is lower than the removal of surface losses. Example field solutions are plotted in Fig.\ref{fig:field}(a).

\begin{figure*}
    \centering
    \includegraphics[width=\linewidth]{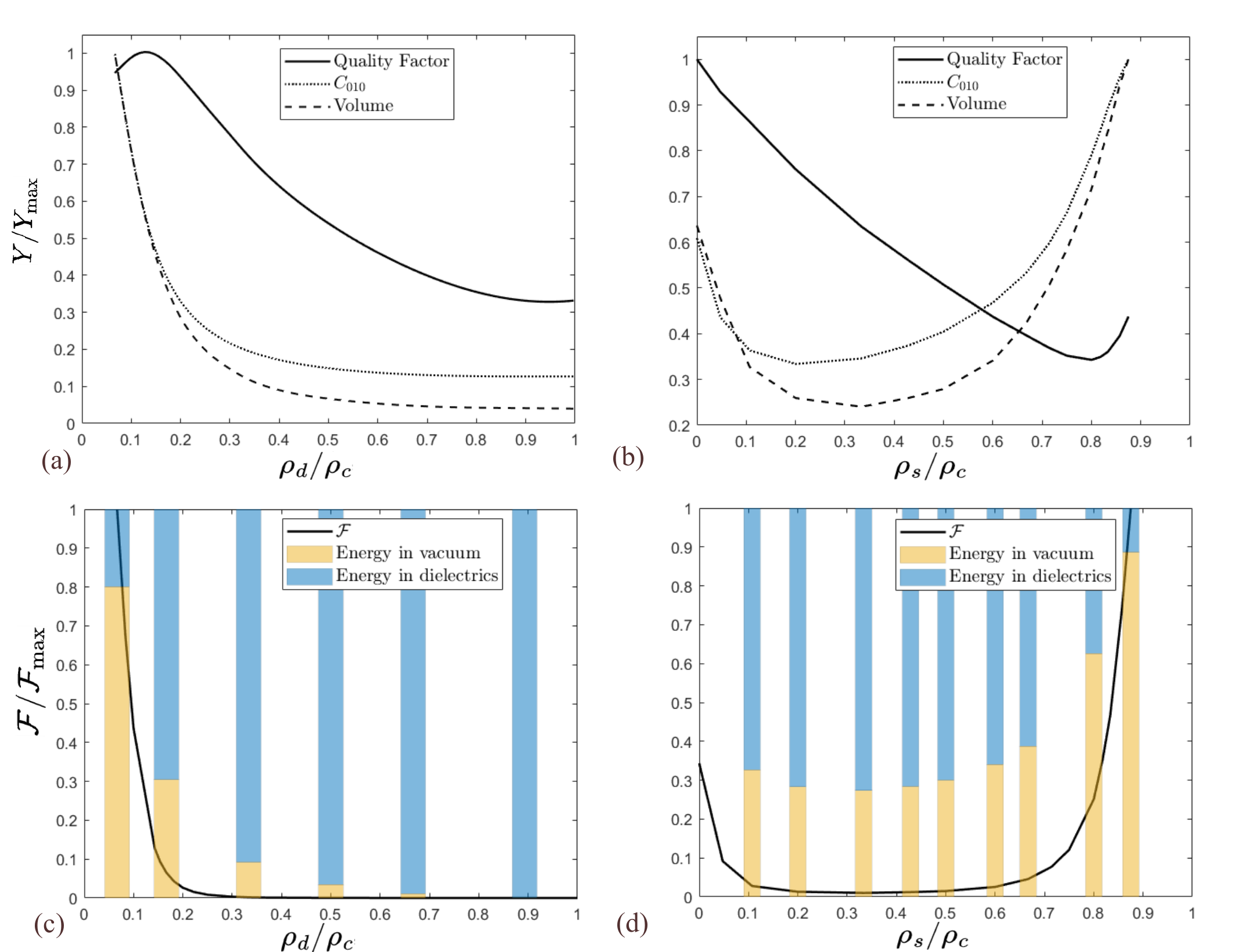} %\captionsetup{justification=raggedright}
    \caption{(a),(b): To select the optimal radial aspect ratio, $Q_0$, $C$, $V$ is plotted against the radial aspect ratio, $\rho_d/\rho_c\ (\rho_s/\rho_c)$, for the dielectric tuning rod cavity (the dielectric shell cavity) while maintaining a constant resonant frequency at 5.8 GHz (the specific frequency choice does not affect the result). The data is cut off before a ratio of 0.1 because it is the minimum ratio for tuning as discussed in Sec.\ref{subsec:radialAR}. (c),(d): The figure of merit $\mathcal{F}$ as a function of $\rho_d/\rho_c\ (\rho_s/\rho_c)$ for the dielectric tuning rod cavity (the dielectric shell cavity) while maintaining a constant resonant frequency at 5.8 GHz. The data is cut off after a ratio of around 0.9 to accommodate for the shell thickness. The energy ratio is over-plotted to show that the less energy resides in the dielectric, the better the cavity performs. To reflect the relative changes, all of the quantities are normalized such that the maximum is 1.}
    \label{fig:AR}
\end{figure*}

\subsubsection{\label{DSC}Dielectric Shell Cavity}
Another application of dielectrics involves reducing power loss at the cavity wall; One such example is surrounding the wall with a dielectric shell, consequently leading to an increase in the Q factor. This model is shown in Fig.\ref{fig:cavity_geometry}(b), where a dielectric shell with thickness $t$ and inner radius $\rho_s$ is located concentrically with the cavity of radius $\rho_c$. It should be noted that the tuning aspect of this design will not be addressed, as it entails developing specialized tuning mechanisms that fall beyond the scope of this study. For the $\text{TM}_{010}$ mode, the electric field in the longitudinal direction is given by

\begin{equation}
E(\rho) =
\begin{cases}
    A_sJ_0(k_0\rho) & 0<\rho \leq \rho_s \\
B_sJ_0(\sqrt{\epsilon_d}k_0\rho)+C_sN_0(\sqrt{\epsilon_d}k_0\rho)&\rho_s<\rho \leq \rho_s +t\\
    D_sJ_0(k_0\rho)+E_sN_0(k_0\rho)&\rho_s +t<\rho \leq \rho_c 
\end{cases}
\label{eq:E_field_ds}
\end{equation}

\begin{comment}
 \begin{eqnarray}
\label{eq:E_field_ds}
\begin{array}{lcr}
    E_{s1}(\rho) = A_sJ_0(k_0\rho) & \mbox{}
    & 0<\rho \leq \rho_s \\
    E_{s2}(\rho) = B_sJ_0(\sqrt{\epsilon_d}k_0\rho)+C_sN_0(\sqrt{\epsilon_d}k_0\rho)&\mbox{}&\rho_s<\rho \leq \rho_s +t\\
    E_{s3}(\rho) = D_sJ_0(k_0\rho)+E_sN_0(k_0\rho)&\mbox{}&\rho_s +t<\rho \leq \rho_c \nonumber
\end{array}
\end{eqnarray}    
\end{comment}
\noindent where $A_s$ through $E_s$  are coefficients to be solved using the matching conditions at the boundary, which are given by
\begin{flalign}
E_{s1}(\rho_{s})=E_{s2}(\rho_{s}),\  E_{s2}(\rho_{s}+t) &=E_{s3}(\rho_{s}+t) \nonumber\\
\frac{\partial E_{s1}}{\partial \rho}|_{\rho=\rho_{s}}=\frac{\partial E_{s2}}{\partial \rho}|_{\rho=\rho_{s}}\nonumber,\ 
\frac{\partial E_{s2}}{\partial \rho}|_{\rho=\rho_{s}+t}&=\frac{\partial E_{s3}}{\partial \rho}|_{\rho=\rho_{s}+t}\nonumber\ \\
E_{s3}(\rho_{s}+t)&=0\nonumber
\end{flalign}

Similar to the dielectric rod cavity, the power loss through the dielectric is calculated using Eq.\ref{eq:P_loss_d} and the shell is again made of sapphire. Example field solutions are plotted in Fig.\ref{fig:field}(b). 

\section{\label{sec:results}Results}

\subsection{\label{subsec:radialAR}Optimal Radial Aspect Ratio}
Because each of the cases described in Sec.\ref{sec:case_studies} has two degrees of freedom in their geometry, size of dielectric element and cavity radii, a given resonant frequency can be achieved with multiple configurations. To ensure a fair comparison between cases, only the optimal geometry, defined by the aspect ratio between the radii ($\rho_m/\rho_c$, $\rho_d/\rho_c$, and $\rho_s/\rho_c$), for each case is used.

To find the optimal aspect ratio of each configuration, we start by maximizing each component of the figure of merit $\mathcal{F}$ defined in Eq.\ref{eq:FOM_scanrate}. For the metal tuning rod cavity, this is straightforward as only the volume depends on the aspect ratio and $\mathcal{F}$, increases as $\rho_m/\rho_c$. For the same frequency, one easily finds that $\rho_c - \rho_m$ is a constant, so the effective volume for a metal rod cavity is proportional to  

\begin{equation}
    V_{\text{eff}} \propto \frac{1+\rho_m/\rho_c}{1-\rho_m/\rho_c}
\end{equation}
where $\frac{1+\rho_m/\rho_c}{1-\rho_m/\rho_c}$ monotonically increases in the range of $0 \le\rho_m/\rho_c<1$. However, the ratio is limited by the need to maintain fine-scale tuning resolution to search for axions of various masses. A large metal rod that is too close to the cavity wall decreases the tuning resolution, leaving insufficient overlap between each spectrum. In addition, the ratio is also limited by the physical scale of the experiment, such as the bore size of the magnet, which constrains the overall size of $\rho_c$. we conservatively set $\rho_m/\rho_c = 1/2$ to match the relative size of the HAYSTAC experiment\cite{HAYSTAC_2023MJ}.

\begin{figure}[ht]
\includegraphics[width=\linewidth]{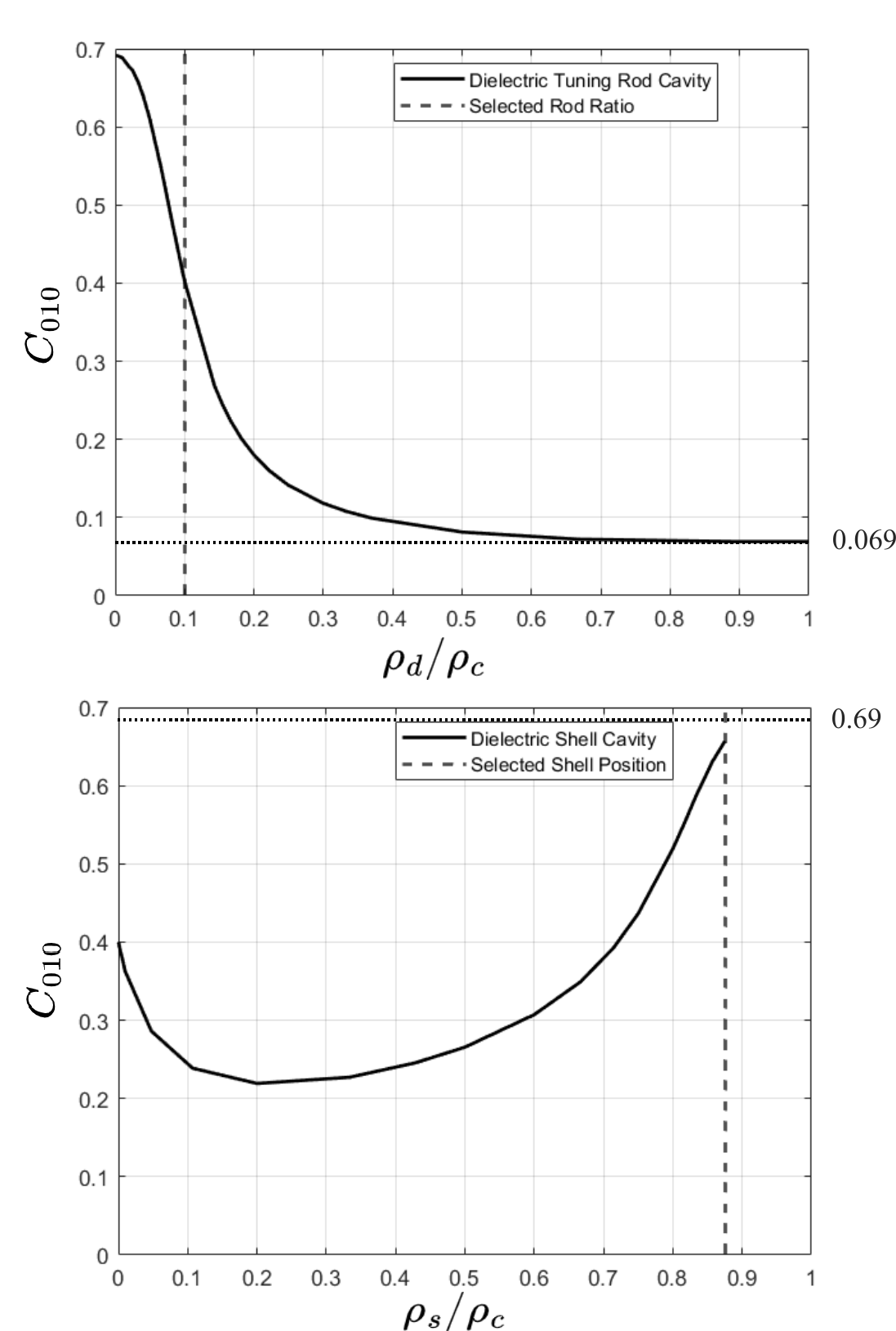}
%\captionsetup{justification=raggedright}
\caption{\label{fig:formfactor_d} (top): The form factor $C_{010}$ of the dielectric rod cavity as a function of the radial aspect ratio. As the rod diameter $\rho_d$ gets smaller, the $C_{010}$ approaches 0.69 for an empty cylindrical cavity. As the rod gets larger, the $C_{010}$ approaches 0.069 for a uniformly-filled dielectric cavity. (bottom): The form factor $C_{010}$ of the dielectric shell cavity as a function of the radial aspect ratio. As the shell moves closer to the wall, $C_{010}$ approaches 0.69 for an empty cavity. For both cases, the vertical dashed line corresponds to the selected aspect ratio for the study, resulting in $C_{010} \sim 0.4$ and $C_{010} \sim 0.66$ for the dielectric rod and dielectric shell respectively. The data for the dielectric shell cavity is cut off after a ratio of around 0.9 to accommodate for the shell thickness.}
\end{figure}

The optimal aspect ratio for the dielectric tuning rod cavity can also be found by minimizing the amount of energy inside dielectrics as suggested in Sec.\ref{subsec:Dielectrics in Resonant Cavities}. Therefore, for the same detection frequency, a smaller rod-cavity ratio is preferred. To verify, we compute the field solution of Eq.\ref{eq:E_field_d} by fixing the resonant frequency and varying $\rho_d/\rho_c$. For the same resonant frequency, $Q_0$, $C_{010}$, $V$, and the dielectric energy ratio are computed as a function of $\rho_d/\rho_c$. The dielectric energy ratio is defined as the ratio of the field energy in the dielectric to the total energy.
\begin{flalign}
    U_{\text{ratio}} &= \frac{U_d}{U_{\text{total}}}\nonumber\\
              &= \frac{\int_{V_d} \epsilon_0 \epsilon_d |\textbf{E}_d|^2 \,d^3x}{\int_{V_d} \epsilon_0 \epsilon_d |\textbf{E}_d|^2 \,d^3x + \int_{V_v} \epsilon_0 |\textbf{E}_v|^2 \,d^3x} 
\end{flalign}
where $V_d$, $\textbf{E}_d$ are the volume and the E-field inside the dielectric, and  $V_v$, $\textbf{E}_v$ are the volume and the E-field of the rest of the cavity, assumed to be vacuum. The results are shown in Fig.\ref{fig:AR} and \ref{fig:formfactor_d}. As the magnetic field amplitude peaks at the dielectric boundary and diminishes towards the cavity wall, employing dielectrics to divert the field away from the wall can help minimize surface loss. An optimal $Q_0$ therefore exists where the rod radius is large enough to draw the field inside and suppress the field outside, but small enough to separate the large magnetic field inside the dielectric from the cavity wall. It can be seen from Fig.\ref{fig:AR}(a) that the aspect ratio that gives the optimal $Q_0$ is $\rho_d/\rho_c\sim1/6$. On the other hand, the volume and form factor $C_{010}$ do not have a natural optimum as they increase arbitrarily with less dielectric material as shown in Fig.\ref{fig:AR}(b). However, because the introduction of the dielectric is meant to tune the cavity's mode frequency over a range of possible axion masses, a practical limit on the rod size is imposed from the desired tuning range. To achieve a balance between the dynamic tuning range and $\mathcal{F}$, $\rho_d/\rho_c$ should be at least 1/10, which yields a tuning range on the order of 100 MHz for a GHz frequency search and a higher $\mathcal{F}$ according to Fig.\ref{fig:AR}(b). 

Similarly, the optimal aspect ratio of a dielectric shell cavity is directly related to its dielectric energy ratio. As shown in Figs.\ref{fig:AR}(d) and \ref{fig:formfactor_d}, the closer the shell is to the cavity wall, the less energy is contained in the dielectric, resulting in better cavity performance. It is also important to keep the shell's thickness small to minimize the energy inside. Because this design is meant to optimize the $Q_0$, we will not worry about the tunability for this study. Instead, we choose the thickness ratio to be $t/\rho_c = 1/10$, and the position of the shell-to-cavity ratio to be $\rho_s/\rho_c = 7/8$ according to Fig.\ref{fig:AR}(d), which is close enough to the wall but far enough to accommodate the shell's thickness.

\subsection{\label{subsec:performance}Performance Comparison}

\begin{figure}[ht]
\includegraphics[width=\linewidth]{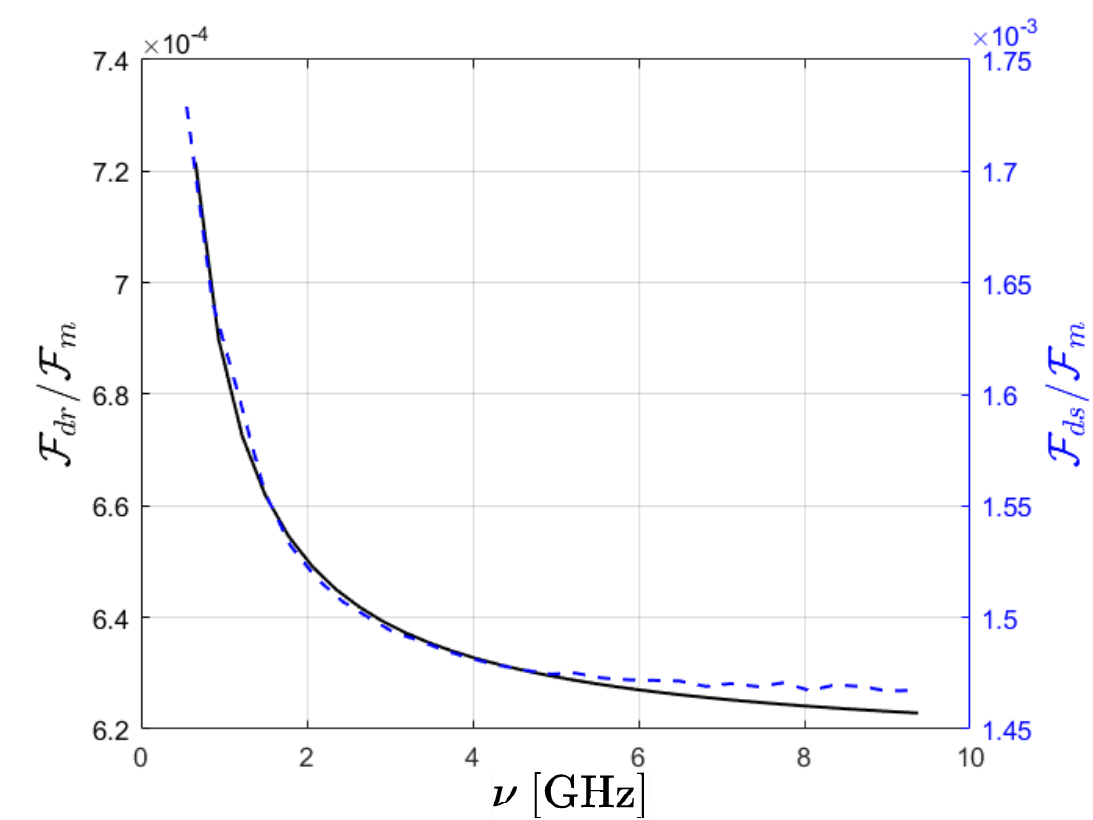}
%\captionsetup{justification=raggedright}
\caption{\label{fig:FOM}The ratio of figure of merits as a function of resonant frequency, shown for both the dielectric rod (solid black) and dielectric shell (dashed blue) relative to the equivalent metal rod at the same frequency. The optimal ratios selected in Sec.\ref{subsec:radialAR} are used when comparing performances ($\rho_m/\rho_c = 1/2$, $\rho_d/\rho_c = 1/10$, $\rho_s/\rho_c = 7/8$). The metal tuning rod cavity constantly outperforms both dielectric configurations between $\sim 0.5$ to 9.5 GHz. The small fluctuations in the plots are the result of rounding errors when calculating the matching conditions.}
\end{figure}     

\begin{figure}[ht]
\includegraphics[width=\linewidth]{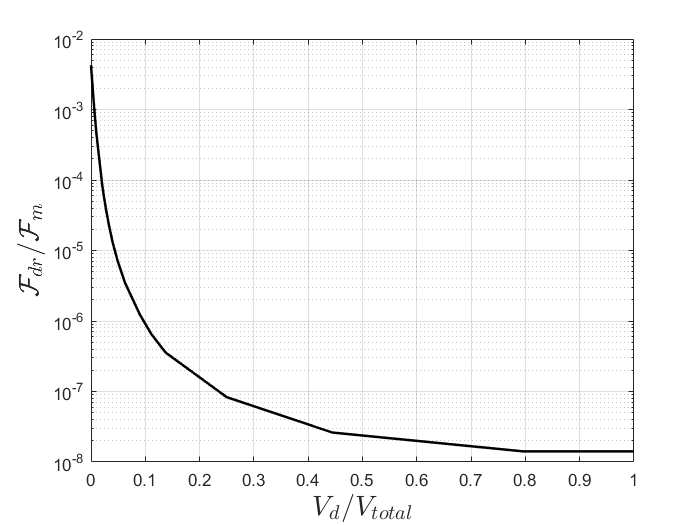}
%\captionsetup{justification=raggedright}
\caption{\label{fig:v_ratio} At the same resonance frequency of 5.8 GHz, the ratio of figure of merit (dielectric rod cavity over metal rod cavity) as a function of dielectric volume. In the absence of dielectrics, an empty cavity with a vacuum has a smaller size than a metal rod cavity at the same resonant frequency, which explains the initial difference. As the proportion of dielectrics increases, the cavity performance deteriorates.}
\end{figure}

Using the optimal aspect ratio selected in Sec.\ref{subsec:radialAR}, we compute $\mathcal{F}$ for all three cases. While not explicitly shown in Eq.\ref{eq:FOM_scanrate}, $\mathcal{F}(\nu)$ scales with frequency roughly as $\nu^{-20/3}$ due to two effects. First, since the resonant TM mode is related to the cavity's radius, the volume of the cavity scales as $\nu^{-3}$ for a fixed aspect ratio. Second, at the temperatures needed to minimize thermal noise, axion detectors are operated in the anomalous skin effects regime\cite{skin_depth_pippard}, where $Q_0 \propto \nu^{-2/3}$. Thus, the resonant frequency should be kept the same when evaluating cavity configurations. We choose to evaluate in a frequency range between $\sim 0.5$ to 9.5 GHz because other non-resonant cavities detection schemes are generally preferred beyond this range\cite {Sikivie_2014, Lamoreaux_2013, Plasmon}. The resulting figure of merit is plotted in Fig.\ref{fig:FOM}. It shows that cavities with dielectrics underperform the metal rod cavity by as much as a factor of $\sim$1600 in scan rate. The main effect is the drop of $\sim$530 from the volume, with the form factor and quality factor accounting for the remaining factor. As discussed in Sec.\ref{subsec:Dielectrics in Resonant Cavities}, the energy per unit length inside a cavity is roughly constant for the same axion frequency. The high permittivity causes the energy to concentrate in one place, thus degrading the performance of the cavity with volume reduction as the leading effect, followed by a decrease in form factor. In Fig.\ref{fig:v_ratio}, the $\mathcal{F}$ ratio of the dielectric rod to the metal rod cavity is plotted against the volume ratio of dielectrics to the total volume. This further reinforces the notion that an increased proportion of dielectrics leads to a decline in cavity performance.

\section{\label{disscussion and conclusion}Discussions and Conclusions}
This work provides an in-depth analysis of the main effects of dielectric materials in a resonant cavity search for axion dark matter. While dielectrics are useful in many cases, such as increasing Q and shifting the frequency of the $\text{TM}_{010}$ to lower frequencies, their use can result in a sub-optimal performance relative to a similar search performed with a cavity devoid of dielectrics. In the simple cases studied in Sec.\ref{sec:case_studies}, this can result in an average of $\sim$1500 reduction in the scan rate over the frequency range of interest for cavity haloscopes. This effect is largely due to the concentration of the field energy in the dielectric, requiring a substantial reduction in volume in order to search at the same frequency achievable with a cavity devoid of dielectrics. This also results in a reduction of the form factor, which can mostly cancel the potential gain in the quality factor from dielectric placement. As such, generally, a cavity devoid of dielectric material is a more favorable configuration for a search using the lowest-order TM mode. However, this ignores the practical limits of lowering the mode frequency of a metal-only cavity which can result in cavity dimensions exceeding the infrastructure used to house the experiment such as the magnet bore size or fridge cooling power. In the case that a metal cavity at the chosen frequency is not possible, the use of dielectrics, while sub-optimal, may offer the exploration of frequencies not accessible otherwise. In the case where there is no preferred frequency target, the best strategy for extending the frequency range of a pre-existing cavity, however, is to move up in frequency with metal elements rather than down in frequency with dielectrics. 

Note that although the models introduced in Sec.\ref{sec:case_studies} have been enhanced with more details, they remain relatively simple. Certain aspects, such as the gap between the dielectric and the cavity caps, are not considered. Nevertheless, this level of simplicity is adequate for providing an order-of-magnitude comparison to demonstrate the primary effect. To further simplify the problem, the tuning of the rod has not been fully accounted for. However, given the limited tuning range of the dielectric rod and the fact that the metal rod is positioned in the middle, representing the worst-case scenario, this comparison remains valid and unbiased. 

As discussed before, in searches utilizing higher-order modes, dielectric materials are used to improve the form factor by suppressing the opposite-phase electric field components. The utilization of dielectrics in higher-order modes is particularly beneficial for high-frequency searches, offering increased volume relative to the lowest mode without a substantial reduction in form factor. While more emphasis is given to the lowest mode in this work, we show that high-order modes have a more stringent longitudinal aspect ratio requirement for limiting the mode crossings, which warrants careful attention in cavity design.

\begin{comment}
It is possible to circumvent the poor volume scaling of dielectric cavities by combining multiple smaller cavities at the cost of added complexity\cite{Hagmann_cavity}. There are ongoing efforts working to reduce the complexity of the multi-cavity operation\cite{Jeong_2018,yang_search_2020}. However, a reduction in form factor due to the dielectrics remains. Furthermore, it is challenging to achieve a wide tuning range with dielectric rods and shells. Sophisticated tuning mechanisms are often required, which adds another layer of complexity. Considering all of the above, when it comes to incorporating large pieces of dielectrics in a haloscope, it becomes a question of how one wants to utilize and maximize the resources. 

  For resonant cavities utilizing higher order $\text{TM}_{0n0}$ modes, the concentration of the energy inside the dielectric may help the cavity performance by reducing the out-of-phase field components\cite{Kim_2020,ORGAN_2018,QUAX_2022}. It is particularly beneficial for high-frequency searches, offering increased volume relative to the lowest mode without a substantial reduction in form factor. While more emphasis is given to the lowest mode in this work, we show that high-order modes have a more stringent longitudinal aspect ratio requirement for limiting the mode crossings (Eq.\ref{eq:AR_requirement_general}). Having dielectrics in these higher-order modes may potentially exacerbate mode crowding, resulting in degraded axion sensitivity in those regions of the frequency space.   
\end{comment}

\section*{Acknowledgements}
The authors thank Samantha Lewis, Sumita Ghosh and Eleanor Graham for helpful discussions and comments on the manuscript. This work is supported by the National Science Foundation under Grant No. PHY-2011357. Michael. J. Jewell and Reina H. Maruyama are also supported in part by the Department of Energy under Grant No. DE-AC02-07CH11359.

\begin{comment}
\section*{AUTHOR DECLARATIONS}
\subsection*{Conflict of Interest}
The authors have no conflicts to disclose.

\section*{DATA AVAILABILITY}
The data that support the findings of this study are available from the corresponding author upon reasonable request.

\section*{Author Contributions}
\end{comment}

\appendix
\section{\label{appen:LAR}longitudinal aspect ratio}
In this appendix, we derive the optimal longitudinal aspect ratio of the cavity which is applied throughout our study. The longitudinal aspect ratio is defined as the ratio of cavity height L over the radius of the confinement area, which is usually the cavity radius $\rho$ for a cylindrical cavity. A certain longitudinal aspect ratio needs to be respected because a higher aspect ratio increases the risk of mode crossings. It suggests that extending the height of the cavity arbitrarily is not an efficient strategy for maximizing the detection volume. 

Mode crossings occur when the TM mode of interest is at a frequency where it becomes degenerate with a TE or TEM mode, resulting in a loss of sensitivity to the axion signal. The mixing of TM modes with other modes degrades the $Q$ of the mode of interest and leaves gaps in the cavity's scan range. For example, in the HAYSTAC experiment, as much as $15\%$ of the available frequency range contains significant mode mixing \cite{rapidis_characterization_2019}. The longitudinal symmetry breaking within the cavities, such as gaps at the rod ends and tilt of the rods, is responsible for the mode crossings\cite{mode_crossing_stern}. In practice, perfect longitudinal symmetry cannot be achieved due to machining and assembly tolerances, which makes mode crossings inevitable. However, we can minimize them by limiting the density of the intruder modes\cite{Hagmann_cavity}.  

Consider the number of TE modes as a function of the wave vector $k$, $N(k)$. In the longitudinal direction, the mode density is $dN_z/dk_z = L/\pi$. In the transverse direction, using the approximation of the Bessel functions of the first kind, we obtain the density of TM or TE modes

\begin{equation}
    \frac{dN_t}{dk_t} \approx \frac{2\rho^2k_t}{\pi}
\end{equation}
Integrating $N_t$ and $N_z$ over k, with constraint $k^2 = k_t^2 + k_z^2$, one finds the total number of modes to be

\begin{equation}
    N(k) = \frac{2\rho^2Lk^3}{3\pi^2}
\end{equation}
Recasting $N(k)$ as $N(\nu)$ with $k = 2\pi \nu/c$ (set $c=1$) and taking the derivative respective to $\nu$, we now obtain the TE mode density as a function of frequency $\nu$

\begin{equation}
    n(\nu) = \frac{dN}{dv} \approx 16\pi \rho^2L\nu^2
    \label{eq:mode_density}
\end{equation}

We now define excessive mode crossings as more than $15\%$ of the total cavity tuning range. For example, a haloscope experiment such as the HAYSTAC has a tuning range of about 2 GHz, and $15\%$ of that corresponds to one major mode crossing every 300 MHz \footnote{In practice,  the interval of mode crossings given by Eq.\ref{eq:AR_requirement_app} will certainly be worse because of the longitudinal symmetry breaking, such as imperfect alignment of the rods, machining flatness tolerance, etc.}. Taking a cavity with a loaded quality factor of $Q_L \sim 5000$ and a bandwidth of $\Delta\nu\sim 1$MHz as an example\cite{HAYSTAC_2023MJ}, this requires tuning about 300 times the cavity bandwidth before hitting a mode crossing, which is a reasonable goal for a practical experiment that wants to cover a wide range. This can be written as
\begin{equation}
     n(\nu)(300 \Delta\nu) = n(\nu)(300 \frac{\nu}{Q_L}) \lesssim 1 
    \label{eq:ARcondition}
\end{equation}
Near the resonant frequency of the $\text{TM}_{0n0}$ mode, $\nu \approx X_{0n}/(2\pi \rho)$ , where $X_{0n}$ is the $n$th zero of the Bessel function $J_0(x)$ with $X_{01}\approx2.4048$, $X_{02}\approx5.5201$, $X_{03}\approx8.6537$, etc. Using Eq.\ref{eq:mode_density} and \ref{eq:ARcondition}, we found for an experiment using TM$_{010}$, the aspect ratio requirement for limiting mode crossings is
\begin{equation}
    \frac{L}{\rho} \lesssim 5, \ \text{for}\ n=1
    \label{eq:AR_requirement_app}
\end{equation}
Similarly for higher-order mode searches, assuming the cavity can tune more than 300 times the cavity bandwidth before hitting a mode crossing, the longitudinal aspect ratio can be expressed in a more general format 
\begin{equation}
\frac{L}{\rho} \lesssim \frac{\pi^2 Q_L}{600X_{0n}^3}.\\
\label{eq:AR_requirement_general_app}
\end{equation}
While neither Eq.\ref{eq:AR_requirement_app} nor Eq.\ref{eq:AR_requirement_general_app} is intended to serve as strict requirements, in the context of this paper, we treat them as approximate guidelines for cavity designs.

%The interference of spurious modes will be even %worse in cavities conducting higher-order mode %searches, which has been observed in %\cite{QUAX_2022}.  

\section{\label{appen:circuit}Cavity Equivalent Circuit Model}

\begin{figure}[ht]
\includegraphics[width=\linewidth]{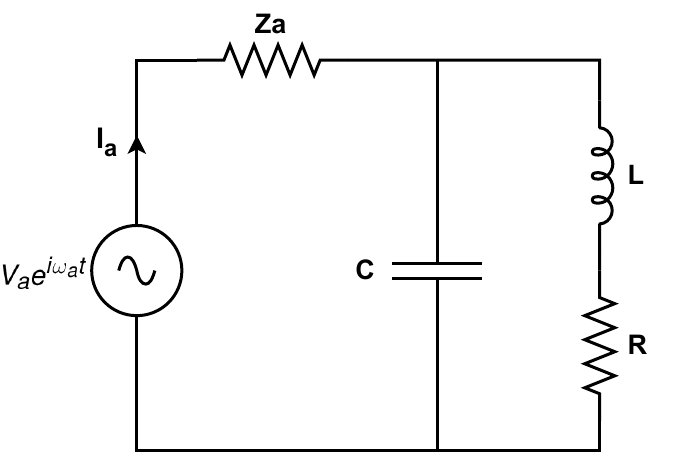}
\caption{\label{fig:circuit} Equivalent circuit of a resonant mode of a cavity. The axion source is represented by an AC voltage source with impedance $Z_a$.}
\end{figure}

 The impedance-matching approach is an alternative to the energy approach shown in Sec.\ref{subsec:Dielectrics in Resonant Cavities} for understanding the effect of dielectrics in a cavity for axion searches. To illustrate impedance-matching, it is convenient to represent the properties of a cavity by its equivalent RLC circuit as shown in Fig.\ref{fig:circuit}. The inductor $L$ and the capacitor $C$ represent the conduction current path and the displacement current path of a cavity respectively\footnote{Please be aware that the symbols $L$ and $C$ used throughout this appendix have different meanings compared to their usage in the main text.}. The resistor $R$ models the loss of the cavity, including resistive loss of the cavity wall, dielectric loss, leakage through antenna holes, etc. The axion source at $\omega_a$ is represented as an AC voltage source $V_ae^{i\omega_a t}$ and provides an ideal current $I_a$. The axion source impedance is given by $Z_a$, which satisfies $\text{Re}(Z_a) \gg R$ because of the small axion-photon coupling and it also prevents back conversion of the axion. The axion conversion power is therefore given by 
\begin{equation}
    P_{ax} = I_a^2 Re(Z_c)
    \label{eq:power_circuit}
\end{equation}
where $Z_c$ is the equivalent complex impedance of the cavity. A simple analysis of the circuit shows that the susceptance is $Y_c=1/Z_c$ is given by 
\begin{equation}
    Y_c = \frac{1}{R+i\omega L}+i\omega C.
\end{equation}
The resonance occurs when the imaginary part goes to zero,
\begin{equation}
\omega_a=\sqrt{\frac{1}{LC}-\frac{R^2}{L}}\approx \frac{1}{\sqrt{LC}}
\end{equation}
and the approximation is valid with a high $Q$, hence low loss inductance, with $R \ll \omega_a L$. 
The real part of the susceptance is then 
\begin{equation}
Y_c=\frac{R}{R^2+(\omega_aL)^2}\approx \frac{R}{(\omega_aL)^2}
\end{equation}
or
\begin{equation}
    Z_c \approx \frac{1}{R}\frac{L}{C}.
\end{equation}

For a fixed axion frequency, $LC$ is constant.
Inserting a dielectric with relative permittivity $\epsilon_d$ increases to $\epsilon_d C=C'$ therefore we must reduce $L$, as $LC=L'C'=(\epsilon C) (L/\epsilon)=C$. 
We then obtain $Z_c \propto \epsilon_d^{-2}$. From Eq.\ref{eq:power_circuit} and taking into account the reduction in electric field from the dielectric material, the axion conversion power is $P_{ax} \propto \epsilon_d^{-3}$, which matches the result from Sec.\ref{subsec:Dielectrics in Resonant Cavities}. (We could have also reduced the cross-sectional area of the capacitor to increase the resonance frequency, which reduces the volume by a factor of $\epsilon$, resulting in $\epsilon_d^2$ reduction in conversion power; however keeping the inductance constant does not model the change in cavity properties by introducing the dielectric.) 

When $L$ is reduced to keep the frequency fixed, in the above circuit analysis we left out an important effect.  As most of $R$ is due to the surface resistance of the inductance, if we reduce $L$ by removing turns from a coil, noting $L\propto n^2$ where $n$ is the number of turns, we can surmise that $R$ scales as $n$, the length of wire in the coil.
Because $L$ must be reduced by a factor $1/\epsilon_d$, then $R\rightarrow R/\epsilon_d$. Therefore,
\begin{equation}
Z_c\rightarrow \frac{\sqrt{\epsilon_d}}{R}\frac{L}{\epsilon_d^2C}\propto \epsilon_d^{-3/2}
\end{equation}
Including the reduction in the electric field, 
$P_{ax} \propto \epsilon_d^{-5/2}$ which now does not correspond to the result in Sec.\ref{subsec:Dielectrics in Resonant Cavities}.

The circuit model provides a physical understanding of the origin of the reduction in $\mathcal{F}$ caused by the presence of the dielectric material in the cavity. In essence, that presence decreases the efficiency of the impedance matching of a cavity to axions by increasing $C$, reducing $L$, but is compensated to some degree by a decrease in $R$ that comes with reducing $L$.  Though there is no exact mapping between the lumped elements and the physical cavity, the results are nonetheless compelling.

\bibliography{aipsamp}

\begin{thebibliography}{10}

\bibitem{peccei1977CP}
R.~D. Peccei and H.~R. Quinn.
\newblock {$CP$} conservation in the presence of pseudoparticles.
\newblock {\em Phys. Rev. Lett.}, 38:1440, Jun 1977.

\bibitem{peccei1977CP2}
R.~D. Peccei and Helen~R. Quinn.
\newblock Constraints imposed by ${CP}$ conservation in the presence of
  pseudoparticles.
\newblock {\em Phys. Rev. D}, 16:1791--1797, Sep 1977.

\bibitem{axionDM_1983}
J.~Ipser and P.~Sikivie.
\newblock Can galactic halos be made of axions?
\newblock {\em Phys. Rev. Lett.}, 50:925--927, Mar 1983.

\bibitem{Sikivie:1983ip_halotheory}
P.~Sikivie.
\newblock {Experimental Tests of the Invisible Axion}.
\newblock {\em Phys. Rev. Lett.}, 51:1415--1417, 1983.
\newblock [Erratum: Phys.Rev.Lett. 52, 695 (1984)].

\bibitem{morris_1984}
D~E Morris.
\newblock Electromagnetic detector for relic axions.
\newblock {\em Lawrence Berkeley Laboratory Techincal Report}, 5 1984.

\bibitem{CAPP}
Ohjoon Kwon, Doyu Lee, Woohyun Chung, Danho Ahn, HeeSu Byun, Fritz Caspers,
  Hyoungsoon Choi, Jihoon Choi, Yonuk Chong, Hoyong Jeong, Junu Jeong, Jihn~E.
  Kim, Jinsu Kim, \ifmmode \mbox{\c{C}}\else \c{C}\fi{}a\ifmmode
  \breve{g}\else~\u{g}\fi{}lar Kutlu, Jihnhwan Lee, MyeongJae Lee, Soohyung
  Lee, Andrei Matlashov, Seonjeong Oh, Seongtae Park, Sergey Uchaikin, SungWoo
  Youn, and Yannis~K. Semertzidis.
\newblock First results from an axion haloscope at capp around $10.7\text{
  }\text{ }\ensuremath{\mu}\mathrm{eV}$.
\newblock {\em Phys. Rev. Lett.}, 126:191802, May 2021.

\bibitem{Alesini_2020_drod}
D.~Alesini, C.~Braggio, G.~Carugno, N.~Crescini, D.~D'Agostino, D.~Di
  Gioacchino, R.~Di Vora, P.~Falferi, U.~Gambardella, C.~Gatti, G.~Iannone,
  C.~Ligi, A.~Lombardi, G.~Maccarrone, A.~Ortolan, R.~Pengo, C.~Pira,
  A.~Rettaroli, G.~Ruoso, L.~Taffarello, and S.~Tocci.
\newblock High quality factor photonic cavity for dark matter axion searches.
\newblock {\em Review of Scientific Instruments}, 91(9):094701, sep 2020.

\bibitem{QUAX_2022}
R.~Di~Vora, D.~Alesini, C.~Braggio, G.~Carugno, N.~Crescini, D.~D'Agostino,
  D.~Di~Gioacchino, P.~Falferi, U.~Gambardella, C.~Gatti, G.~Iannone, C.~Ligi,
  A.~Lombardi, G.~Maccarrone, A.~Ortolan, R.~Pengo, A.~Rettaroli, G.~Ruoso,
  L.~Taffarello, and S.~Tocci.
\newblock High-$q$ microwave dielectric resonator for axion dark-matter
  haloscopes.
\newblock {\em Phys. Rev. Applied}, 17:054013, May 2022.

\bibitem{QUAX_2021}
D.~Alesini, C.~Braggio, G.~Carugno, N.~Crescini, D.~D' Agostino, D.~Di
  Gioacchino, R.~Di Vora, P.~Falferi, U.~Gambardella, C.~Gatti, G.~Iannone,
  C.~Ligi, A.~Lombardi, G.~Maccarrone, A.~Ortolan, R.~Pengo, C.~Pira,
  A.~Rettaroli, G.~Ruoso, L.~Taffarello, and S.~Tocci.
\newblock Realization of a high quality factor resonator with hollow dielectric
  cylinders for axion searches.
\newblock {\em Nuclear Instruments and Methods in Physics Research Section A:
  Accelerators, Spectrometers, Detectors and Associated Equipment}, 985:164641,
  jan 2021.

\bibitem{ORGAN_2018}
Ben~T. McAllister, Graeme Flower, Lucas~E. Tobar, and Michael~E. Tobar.
\newblock Tunable supermode dielectric resonators for axion dark-matter
  haloscopes.
\newblock {\em Physical Review Applied}, 9(1), jan 2018.

\bibitem{blair1987}
D~G Blair and S~K Jones.
\newblock A high-q sapphire loaded superconducting cavity resonator.
\newblock {\em Journal of Physics D: Applied Physics}, 20(12):1559, dec 1987.

\bibitem{simanovskaia_microwave_2023}
Maria Simanovskaia, Gianpaolo Carosi, and Karl~van Bibber.
\newblock Microwave cavity searches.
\newblock In Derek~F. Jackson~Kimball and Karl van Bibber, editors, {\em The
  Search for Ultralight Bosonic Dark Matter}, pages 123--139. Springer
  International Publishing, 2023.

\bibitem{Kim_2020}
Jinsu Kim, SungWoo Youn, Junu Jeong, Woohyun Chung, Ohjoon Kwon, and Yannis~K
  Semertzidis.
\newblock Exploiting higher-order resonant modes for axion haloscopes.
\newblock {\em Journal of Physics G: Nuclear and Particle Physics},
  47(3):035203, feb 2020.

\bibitem{Chaudhuri_2021}
Saptarshi Chaudhuri.
\newblock Impedance matching to axion dark matter: considerations of the
  photon-electron interaction.
\newblock {\em Journal of Cosmology and Astroparticle Physics}, 2021(12):033,
  dec 2021.

\bibitem{loss_tan_sapphire}
Jerzy Krupka, Krzysztof Derzakowski, Michael Tobar, John Hartnett, and
  Richard~G Geyer.
\newblock Complex permittivity of some ultralow loss dielectric crystals at
  cryogenic temperatures.
\newblock {\em Measurement Science and Technology}, 10(5):387–392, 1999.

\bibitem{HAYSTAC_2023MJ}
M.~J. Jewell et~al.
\newblock {New Results from HAYSTAC's Phase II Operation with a Squeezed State
  Receiver}.
\newblock 1 2023.

\bibitem{skin_depth_pippard}
A.B. Pippard.
\newblock The surface impedance of superconductors and normal metals at high
  frequencies ii. the anomalous skin effect in normal metals.
\newblock {\em Proceedings of the Royal Society of London. Series A.
  Mathematical and Physical Sciences}, 191(1026):385–399, 1947.

\bibitem{Sikivie_2014}
P.~Sikivie, N.~Sullivan, and D.B. Tanner.
\newblock Proposal for axion dark matter detection using an lc circuit.
\newblock {\em Physical Review Letters}, 112(13), mar 2014.

\bibitem{Lamoreaux_2013}
S.~K. Lamoreaux, K.~A. van Bibber, K.~W. Lehnert, and G.~Carosi.
\newblock Analysis of single-photon and linear amplifier detectors for
  microwave cavity dark matter axion searches.
\newblock {\em Physical Review D}, 88(3), aug 2013.

\bibitem{Plasmon}
Matthew Lawson, Alexander~J. Millar, Matteo Pancaldi, Edoardo Vitagliano, and
  Frank Wilczek.
\newblock Tunable axion plasma haloscopes.
\newblock {\em Phys. Rev. Lett.}, 123:141802, Oct 2019.

\bibitem{rapidis_characterization_2019}
Nicholas~M. Rapidis, Samantha~M. Lewis, and Karl~A. van Bibber.
\newblock Characterization of the {HAYSTAC} axion dark matter search cavity
  using microwave measurement and simulation techniques.
\newblock {\em Rev. Sci. Instrum.}, 90(2):024706, 2019.
\newblock \_eprint: 1809.02246.

\bibitem{mode_crossing_stern}
I.~Stern, G.~Carosi, N.S. Sullivan, and D.B. Tanner.
\newblock Avoided mode crossings in cylindrical microwave cavities.
\newblock {\em Phys. Rev. Applied}, 12:044016, Oct 2019.

\bibitem{Hagmann_cavity}
C.~Hagmann, P.~Sikivie, N.~Sullivan, D.~B. Tanner, and S.‐I. Cho.
\newblock Cavity design for a cosmic axion detector.
\newblock {\em Review of Scientific Instruments}, 61(3):1076--1085, 1990.

\end{thebibliography}
\bibliographystyle{unsrt}
\end{document}